\documentclass[12pt,preprint]{aastex}
\begin{document}
\title{A Second Kelvin-Helmholtz Timescale of Post Helium-Flash Evolution}

\author{Andrew Gould}
\affil{Department of Astronomy, Ohio State University,
140 W.\ 18th Ave., Columbus, OH 43210, USA; 
gould@astronomy.ohio-state.edu}

\begin{abstract}

I show that after the ``helium flash'' abruptly ends its first
ascent red giant evolution, a solar-type star is powered primarily
by gravitational contraction of its helium core, rather than
by nuclear fusion.  Because this energy is released in the core
rather than the envelope, the overall structure of the star,
and so its luminosity, is driven toward that of a red clump
star from its initial position at the tip of the red giant branch (TRGB).
This occurs on a first (and well recognized) Kelvin-Helmholtz timescale
$t_{\rm KH,1}\sim E_{\rm env}/L_{\rm TRGB}\sim 10^4\,{\rm yr}$, 
where $E_{\rm env}$ is the thermal energy stored in the envelope and
$L_{\rm TRGB}$ is the luminosity at the TRGB.  However, once the
star assumes the approximate structure of a clump star, it remains
powered primarily by contraction for a {\it second} Kelvin-Helmholtz timescale
$t_{\rm KH,2}\sim E_{\rm core}/L_{\rm clump}\sim 10^6\,{\rm yr}$, 
where $E_{\rm core}$ is the thermal energy stored in the core and
$L_{\rm clump}$ is the luminosity of a clump star.  It is this
second Kelvin-Helmholtz timescale that determines the overall pace of
the moderately violent processes by which the star returns to 
nuclear-power generation as a full-fledged clump star.  The reservoir
of gravitational energy acts as ultimate regulator, providing whatever
supplemental energy is needed to power $L_{\rm clump}$ and 
occasionally absorbing the large momentary excesses from helium
mini-flashes.  As this reservoir is gradually exhausted, helium
fusion approaches the level of steady-state clump stars.

\end{abstract}

\keywords{stars: fundamental parameters}

\section{{Introduction}
\label{sec:intro}}

The transition from hydrogen-shell burning first ascent
giants to helium-core burning clump stars is triggered by a ``helium flash''.
The underlying physics of the helium flash is so simple and so
compelling that confidence in this conclusion is universal 
despite (until very recently) a complete absence of physical evidence.
Shell burning gradually deposits helium ash in the core of the giant,
which forms a degenerate ``white dwarf'' with mass-radius relation
$R\propto M^{-1/3}$.  Just above the core, the hydrogen is
compelled by the virial theorem to maintain a temperature satisfying
$(3/2)KT = (1/2)GM\mu_H/R$ where $\mu_H$ is the mean molecular weight
of the hydrogen gas. Hence $T\propto M^{4/3}$.  The helium core
is slaved to this thermostat.  Since the core is supported by
degeneracy pressure, it initially hardly expands as its temperature
rises.  Once the temperature rises to the point that heat is
being generated faster than it can escape, a runaway sets in, since
helium fusion scales $\sim T^{40}$.  Eventually, enough heat is released
to lift the degeneracy, which shuts off the runaway via both
adiabatic cooling and reduced density $\rho$, since helium fusion
scales $\sim\rho^3$.  ``Eventually'', the core becomes thermally supported
and burns helium in a continuous manner, at which time it becomes
a clump star.  Since stars at the tip of the red giant branch (TRGB)
and the clump are observed and are very well modeled according to
the above picture, there can hardly be any doubt that the helium
flash occurs.

Nevertheless, despite tremendous advances in modeling the 
post-helium-flash transition toward the clump,
the underlying physics of this transition remains obscured
beneath a welter of calculated details.  It is known, for example
that the transition takes of order 2 Myr, that during
most of this transition, the star appears in most respects as
a ``normal'' clump star, that it experiences a series of helium
mini-flashes.  This picture has been gradually developed over 
four decades, including particularly work by \citet{thomas67},
\citet{iben84}, and \citet{mocak08}.  Most recently, \citet{bildsten12}
have extracted detailed predictions for acoustic signatures from
their models of this 2 Myr period of transition.  Very strikingly,
three stars observed by the {\it Kepler} and {\it CoRoT} missions
exhibit the predicted signatures, bringing the helium-flash
triggered transition into the realm of observational astrophysics
for the first time.

However, despite these advances, the basic reason for the $\sim$ Myr
timescale has never been given.  And related to this, the main
energy source powering the star's luminosity has never been specified,
although of course this energy source is fully encoded in the models.
Here I resolve both issues.

\section{{Second Kelvin-Helmholtz Timescale}
\label{sec:kh2}}

It is of course well known that the energy required to transition
from a degenerate core to a helium-burning core at the temperature,
$k T_{\rm He}\sim 9\,$KeV, required to burn helium is given by
\begin{equation}
\Delta E \sim {1\over 2}\biggl( 
{G M_c^2\over R_{\rm degen}} - {G M_c^2\over R_{\rm therm}}\biggr)
= \epsilon_{\rm trans} M_c c^2\sim 100\,{\rm Myr}\times L_\odot
\label{eqn:deltae}
\end{equation}
where $\epsilon_{\rm trans}$ is evaluated by applying the virial theorem
to the hydrogen thermostat before transition and to the core itself
afterward,
\begin{equation}
\epsilon_{\rm trans} \sim 
{3\over2} \biggl(
{k T_{\rm He}\over \mu_{\rm H}c^2}-{k T_{\rm He}\over \mu_{\rm He}c^2}\biggr)
\sim 1.2\times 10^{-5}.
\label{eqn:epstrans}
\end{equation}
Here, $M_c\sim 0.5\,M_\odot$ is the mass of the core at the helium flash,
$\mu_{\rm H}c^2= 0.58\,$GeV is the mean molecular weight of the hydrogen
mixture and $\mu_{\rm He}c^2\sim 1.25\,$GeV is the mean molecular weight
of helium.

I now argue that it is the Kelvin-Helmholtz time derived from this energy,
\begin{equation}
t_{\rm KH,2} = {\Delta E\over L_{\rm clump}} \sim 2\,{\rm Myr}
\label{eqn:kh2}
\end{equation}
that is required for the core to adjust to steady-state helium burning,
independent of the details of the transition.

The first point is that the helium flash must itself provide of order
$\Delta E$ in order to at least partially lift the degeneracy and
so shut down the runaway helium burning.  Most of this energy is injected
on timescales of seconds or minutes, so that it may
appear that in principle the transition could take place very quickly.
However, quick examination of two limiting cases shows that this
cannot be so.

In one limit, the runaway-shutdown is effective after injecting just
a modest fraction of $\Delta E$.  In this case, the rest of the
energy must be supplied by something approximating core helium burning,
which would require a time $t_{\rm KH,2}=\Delta E/L_{\rm clump}$.
In the other limit, the energy injection process overshoots the
minimum energy required to lift the degeneracy.  Such an overshoot
must be less than a factor 2, or the star would blow up (which manifestly
does not happen).  In this case, gravitational contraction of the
over-expanded core will power the star for a time 
$\Delta E/\langle L\rangle$, where $\langle L\rangle$ is the mean
luminosity of the star during this epoch.

This brings me to the second point, which is that in either of the
above two cases, energy generation is taking place in the core,
as it does in clump stars but not TRGB stars.  This implies that, regardless
of the energy source (nuclear fusion or gravitational contraction),
the star as a whole will be driven toward the structure of a clump
star, and hence toward a luminosity $L_{\rm clump}$.  This will
take place on a timescale $t_{\rm KH,1} = E_{\rm env}/L_{\rm TRGB}$
which is much shorter than $t_{\rm KH,2}$.  Hence 
$\langle L\rangle\sim L_{\rm clump}$, and thus the second
case also implies a timescale $t_{\rm KH,2}=\Delta E/L_{\rm clump}$.
Thus, in both limiting cases, the timescale of transition from the
TRGB to the clump is $t_{\rm KH,2}$.

The actual helium flash is likely to be a combination of these
two extreme cases rather than intermediate between them.  The
flash is triggered off center because of neutrino cooling of the
deep core.  Hence, the outer part of the core is likely to receive
more than enough energy to lift the degeneracy and the inner core
too little.  Both zones play important roles in the overall behavior
of the star as it makes its transition.

The outer-zone overshoot implies that there is enough gravitational-contraction
energy to power the star for $\sim 1\,$Myr at $L_{\rm clump}$.
Ordinarily, such an ``external'' energy source would tend to shut
down nuclear fusion.  That is, the nuclear burning is normally regulated
by a balance between pressure and gravity: energy loss due to luminosity
would lead to collapse of the star in the absence of energy generation,
and this drives fusion to a rate commensurate with this loss.  But
if the energy is provided by gravitational contraction 
(as in a pre-main-sequence star) the pressure is not driven to the
level required to sustain burning.  Thus, the outer-zone overshoot by itself
would tend to imply that helium fusion should be shut off, or at least
heavily suppressed, during the transition.

However, if the degeneracy is not lifted near the center of the core,
as seems likely from the asymmetry of the flash, then conditions
here more closely approximate those of the pre-flash star.  Since
adiabatic expansion has cooled the core as a whole, this region
is not immediately in contact with a ``thermostat region'' at the
helium-burning temperature $T_{\rm He}$, i.e., the role played by the
hydrogen burning shell at the time of the flash.  

\subsection{{Approach to Steady-State Clump}
\label{sec:steady}}

As contraction exhausts the reservoir of gravitational energy, the
core must replace that source of luminosity with helium fusion.
This fusion must take place near the boundary between the remaining
degenerate inner core and the degeneracy-lifted outer core
because heat transport is much more efficient in the latter
(conduction vs.\ convection), making this the hottest part of
the star.  Fusion will go faster on the inner side of the
boundary because the density is higher while the
temperatures are the same.  

There are only three things that can happen to energy generated 
in the degenerate core: 1) migrate outward and so contribute to 
the luminosity of the star; 2) migrate inward and so trigger
deeper burning: 3) stay in the same region, heating it and so 
accelerating fusion, and thus triggering runaways leading to
additional mini-flashes.

While it is not possible to capture the details of this process
from this sort of general argument, the fact that conduction
moves heat from warmer to colder regions implies that (2)+(3)
must exceed (1).  Since (1) is contributing a substantial
fraction of $L_{\rm clump}$, and since the remaining degenerate
core mass $M_{\rm degen}$ contains only a small fraction,
$(M_{\rm degen}/M_c)^{7/3}$, of the original binding energy, the timescale for
dissolution of the remaining degenerate core is smaller than
$t_{\rm KH,2}$.

Because this timescale is shorter, in practice the dissolution
of the remaining degenerate core
will not take place {\it after} gravitational contraction is exhausted 
but rather simultaneously with it.

\section{{Comparison to Models}
\label{sec:models}}

The above, very general arguments imply that independent of the
details of the process,
\hfil\break\noindent
{1.} The transition should take of order 1--2 Myr, during which
the star should appear externally as a clump star.
\hfil\break\noindent
{2.} The star should, at the beginning, be powered primarily by 
gravitational contraction, with helium fusion heavily suppressed.
Then helium burning should gradually take over as the reservoir of
gravitational energy is exhausted.
\hfil\break\noindent
{3.} Helium fusion should take place primarily in mini-flashes
at the beginning, and then gradually approach the continuous-burning
rate of a clump star toward the end

\citet{bildsten12} have recently published detailed 
models of the transition period.  Their calculations contain vastly more
physics than the above simple arguments, and as a result make
detailed predictions of acoustic signatures that should be observable
in {\it Kepler} and {\it CoRoT} data, and arguably have already been
observed.  This is the first prediction of transition observables and
hence the first real opportunity to test the details of models.
However, independent of whether these details are correct, the above
general points should hold, and they should hold for the 
\citet{bildsten12} models in particular.

All three ``predictions'' are ``confirmed'' by inspection of 
Figure 2 from \citet{bildsten12}\footnote{I have placed these words
in quotation marks because in fact I only formulated these
``predictions'' after seeing the \citet{bildsten12} figure.  My
actual intellectual process was to recognize that the principal
features of this figure flowed from very general considerations,
and did not depend in any way on the details of the calculation.
In this {\it Letter}, I focus on these general considerations,
which I believe could in principle have been elucidated without
reference to specific models.  But the fact remains that they
were not.}.   First, the stellar luminosity
remains fixed at almost exactly $L_{\rm clump}$ during the entire
2 Myr transition (orange line).  Second, the combined power of
the hydrogen and helium fusion (green and blue lines)
remains at a few percent of $L_{\rm clump}$ for most of the first 0.5 Myr,
implying that most
of the energy powering the total luminosity comes from gravitational
contraction.  Third, during this period, the core radius (brown line)
shrinks from roughly 0.1 to 0.075 $R_\odot$, which should release
roughly $12\,{\rm Myr}\, L_\odot$ of energy, about enough to fully power a
clump star for 0.3 Myr.  Finally, the transition is punctuated by 
helium mini-flashes that briefly raise the core luminosity well
above $L_{\rm clump}$, which implies that time-averaged helium-fusion
in the remaining degenerate core
provides most of the luminosity not already generated by gravitational
contraction.

%\begin{equation}
%\label{eqn:}
%\end{equation}

\acknowledgments
I thank Juna Kollmeier, Jennifer Van Saders, Lars Bildsten, and Scott
Gaudi for stimulating discussions.
This work was supported by NSF grant AST 1103471.

%\begin{figure}
%\plotone{cmd.ps}
%\caption{\label{fig:cmd}
%}
%\end{figure}

\end{document}